\documentclass[english]{IEEEtran}
\usepackage[T1]{fontenc}
\usepackage[latin9]{inputenc}
\usepackage{amsmath}
\usepackage{graphicx}
\usepackage{epstopdf}
\usepackage{babel}
\usepackage{cite}
\usepackage{flushend}

\DeclareMathOperator*{\argmax}{arg\,max}

\IEEEoverridecommandlockouts
\IEEEpubid{This paper was submitted to the IEEE SCVT conference 2016} 

\begin{document}

\title{Performance Comparison of Short-Length Error-Correcting Codes}

\author{\IEEEauthorblockN{J.~Van~Wonterghem$^*$, A.~Alloum$\dagger$, J.J.~Boutros$\ddagger$, and M.~Moeneclaey$^*$\\}
\IEEEauthorblockA{\IEEEauthorrefmark{1}Ghent University, 9000 Ghent, Belgium, johannes.vanwonterghem@ugent.be}\\
\IEEEauthorblockA{\IEEEauthorrefmark{2}Nokia Bell Labs, 91620 Nozay, France, amira.alloum@nokia-bell-labs.com}\\
\IEEEauthorblockA{\IEEEauthorrefmark{3}Texas A\&M University, 23874 Doha, Qatar, boutros@tamu.edu
}
}

\maketitle

\begin{abstract}
We compare the performance of short-length linear binary codes on the  binary erasure channel
and the binary-input Gaussian channel. We use a universal decoder that can decode any linear binary block code:
Gaussian-elimination based Maximum-Likelihood decoder on the erasure channel 
and probabilistic Ordered Statistics Decoder on the Gaussian channel. 
As such we compare codes and not decoders. The word error rate versus the channel parameter is found
for LDPC, Reed-Muller, Polar, and BCH codes at length 256 bits. 
BCH codes outperform other codes in absence of cyclic redundancy check. 
Under joint decoding, the concatenation of a cyclic redundancy check makes all codes perform
very close to optimal lower bounds.
\end{abstract}
\section{Introduction}
One of the main goals of Information Theory founded by C.E.~Shannon in 1948
is to make digital communications over a noisy channel. Applications
are found in nowadays technology in fourth and fifth generations of mobile networks, 
in digital video broadcast, in optical communications at the Internet backbone, etc.
Information Theory \cite{Cover2006} predicts the existence of good error-correcting codes 
that are capable of achieving channel capacity. These optimal codes can transmit
at the highest possible information rate given the noise level in the channel.
In the past half century, mathematicians and engineers built many families
of error-correcting codes~\cite{MacWilliams1977},\cite{Blahut2003}, 
to make true (or almost true) the performance predicted
by Information Theory. The latter is a theory for asymptotically long codes.
At asymptotic length, the code analysis is easier~\cite{Richardson2008} (think about the law of large numbers).
Also, in some specific applications such as fiber optic communications
and data storage, huge packets of data are used allowing the application of
capacity achieving codes. In recent fifth generation systems which are currently under construction, 
engineers are interested in short length packets. In parallel, at the theoretical side,
researchers are studying the finite length regime of codes~(e.g.~\cite{Polyanskiy2010}).
Our paper is dedicated to error-correcting codes of short length, typically 256 bits.
Phenomena observed at asymptotic length (100 thousand bits and above) like
channel polarization~\cite{Arikan2009} and threshold saturation~\cite{Kudekar2013} 
do not have a practical effect at these short lengths.

In this work, we compare the performance of short-length linear 
binary codes under equal-complexity identical
decoding conditions based on a universal decoder. Complexity is not the main issue of this paper.
We aim at comparing codes with respect to their performance by means of the best possible decoder
which yields Maximum Likelihood (ML) or near Maximum Likelihood error rates.
Two channels are considered for this performance comparison:
the binary erasure channel (BEC) and the binary input additive white Gaussian noise (BI-AWGN) channel.
The former channel corrupts the transmitted codeword by erasing some of its bits, the latter
adds white Gaussian noise to the observed values corresponding to the codeword bits.
To recover the original information transmitted over the channel, the receiver has to decode the 
corrupted observation. Over the years, many decoding strategies have been developed, often
specific to one family of error-correcting codes \cite{Blahut2003},\cite{Richardson2008},\cite{Ryan2009}. 
For our comparison, we use a universal decoder that can decode any linear binary block code:
Gaussian-elimination based ML decoder on the BEC and probabilistic Ordered Statistics Decoder (OSD)
on the BI-AWGN channel. 
Moreover, the decoders under consideration are also optimal$/$near-optimal
whereas many decoding strategies are sub-optimal, favoring decoding speed over performance.
As a result we compare codes and not decoders.

The paper is structured as follows. Notation and ML decoding on the erasure channel are described in the next section.
Section~\ref{sec_osd} briefly explains OSD decoding and its improvements. 
The list of linear binary block codes considered in this paper is found in Section~\ref{sec_codes}.
Section~\ref{sec_results} includes performance results in terms of word error rate.
Our conclusions on the performance of short-length codes are drawn in the final section.

\IEEEpubidadjcol

\section{Notation and ML decoding on the BEC \label{sec_bec}}
The first scenario we consider is that of ML decoding on the BEC channel. 
At the transmitter, a length-$k$ binary information message $\boldsymbol{b}=\left(b_{1},...,b_{k}\right)$,
consisting of independent and identically distributed (i.i.d.) bits with $P(b_{i}=0)=P(b_{i}=1)=1/2$, is encoded
into a binary coded message $\boldsymbol{c}=(c_{1},...,c_{n})$ of
length $n$ using a linear binary block code $C$. The code $C$ is completely
specified by its $k\times n$ generator matrix $G$~\cite{MacWilliams1977}. 
In systematic form we have $G=[I_{k}|P]$,
where $I_k$ is the $k \times k$ identity and $P$ is a parity matrix defining the code.
The encoding operation can be written as $\boldsymbol{c}=\boldsymbol{b}G=[\boldsymbol{b}~|~\boldsymbol{p}]$
with $\boldsymbol{p}$ the parity bits corresponding to $\boldsymbol{b}$.
This codeword is transmitted over the BEC channel with erasure probability
$\epsilon$ such that at the receiver we get the sequence $\boldsymbol{y}$
with $P(y_{i}=?)=\epsilon$ and $P(y_{i}=c_{i})=1-\epsilon$, for $i=1 \ldots n$,
i.e. the symbol transmitted over the channel is erased with probability $\epsilon$.
At the receiver, ML decision is performed to construct an estimate
$\hat{\boldsymbol{b}}=\argmax P(\boldsymbol{y}|\boldsymbol{b})$
of the originally transmitted $\boldsymbol{b}$, based on the observation
$\boldsymbol{y}$.

Maximum Likelihood decoding of a linear binary code on the erasure channel is equivalent to filling bits, 
as much as possible, among those erased by the channel. 
The verb {\em fill} is equivalent to {\em solve} in this context.
Let $H$ be the $(n-k) \times n$ parity-check matrix of $C$.
The parity-check constraint $H\boldsymbol{c}^t=0$ is true 
for every codeword $\boldsymbol{c}$ in $C$~\cite{MacWilliams1977}.
The linear system $H\boldsymbol{c}^t=0$ is used to fill erasures via Gaussian elimination.
Let $w$ be the erasure weight, i.e. $w$ is the number of erased bits in the transmitted codeword $\boldsymbol{c}$.
Assume that $C$ has parameters $(n,k,d_H)$, where $d_H$ is its minimum Hamming distance. 
For $w \le d_H-1$, all $w$ erasures in any $w$ positions can be filled by an algebraic decoder 
or a ML decoder~\cite{Blahut2003}. For non-trivial binary codes, for $d_H\le w\le n-k$, $n-k$ being the rank of $H$,
algebraic decoding fails because it is bounded by $d_H$
whereas ML decoding based on solving $H\boldsymbol{c}^t=0$ may fill a fraction of the erased bits or all of them.

ML decoding via Gaussian elimination has an affordable complexity, at least in software applications,
for a code length $n$ as high as a thousand bits. The cost of solving $H\boldsymbol{c}^t=0$ is $O(n\times (n-k)^2)$.
Results shown in Section~\ref{sec_results} are obtained for a short length $n=256$. 
\section{OSD decoding on the BI-AWGN \label{sec_osd}}
The transmitter for the BI-AWGN channel is the same as for the BEC channel,
notation is inherited from the previous section.
Before transmission on the BI-AWGN channel, the coded message is mapped to a BPSK symbol sequence
$\boldsymbol{s}\in\{-1,+1\}^{n}$ using the rule $s_{i}=2c_{i}-1$.
This symbol sequence is transmitted over the AWGN channel characterized by its single sided noise spectral density $N_{0}$.
At the output of the channel, we receive $\boldsymbol{r}=\boldsymbol{s}+\boldsymbol{w}$
where $\boldsymbol{w}=(w_{1},...,w_{n})$ is a set of i.i.d. real Gaussian
random variables with zero mean and variance $\sigma^{2}=N_{0}/2$.
Note that the symbols $s_i$ are normalized to unit energy such that the
energy transmitted per information bit equals $E_b = \frac{n}{k}$.

At the receiver soft-decision decoding is performed to construct an
estimate $\hat{\boldsymbol{b}}$ of the originally transmitted information
message $\boldsymbol{b}$. For this estimate, the decoder makes use
of two vectors corresponding to the sign and magnitude of the received
signal $\boldsymbol{r}$:\\
The hard-decision $\boldsymbol{y}=[\boldsymbol{b}_{\text{HD}}|\boldsymbol{p}_{\text{HD}}]$ where 
\[
y_{i}=\begin{cases}
0 & \text{for }r_{i}<0\\
1 & \text{for }r_{i}\ge0
\end{cases}
\]
and the confidence values 
\[
\alpha_{i}=\left|r_{i}\right|, ~~~i=1\ldots n.
\]
To understand that $\alpha_{i}$ is indeed a measure for the confidence of the received
$r_{i}$, it suffices to see that the log-likelihood ratio is 
$\text{\ensuremath{\Lambda}}_{i}=\log\frac{P(c_{i}=0|r_{i})}{P(c_{i}=1|r_{i})}=\frac{2r_{i}}{\sigma^{2}}$
for the considered system.

A hard-decision decoder only
uses the vector $\boldsymbol{y}$ to produce its estimate $\hat{\boldsymbol{b}}$.
The omission of the information contained in the magnitude of $\boldsymbol{r}$
explains why hard-decision decoders perform worse than soft-decision
decoders (page 15 in \cite{Ryan2009}).

\subsection{Soft-decision decoding using the OSD algorithm}
Soft-decision decoding by the receiver is performed using the OSD algorithm,
an efficient most reliable basis (MRB) decoding algorithm
firstly proposed by Dorsch \cite{Dorsch1974}, further developed by Fang and Battail \cite{Fang1987}, 
and later analyzed and revived by Fossorier and Lin~\cite{Fossorier1995}.
In the first step of the algorithm, the received
vector $\boldsymbol{r}$ is sorted in order of descending confidence
and the corresponding permutation $\pi_{1}$ is applied to the generator
matrix $G$, yielding $G'$. Gaussian elimination is now performed
on $G'$ to construct the systematic $\tilde{G}$, note that an additional
permutation $\pi_{2}$ may be necessary. We write $\tilde{\boldsymbol{y}}=\pi_{2}\left(\pi_{1}\left(\boldsymbol{y}\right)\right)=\left[\tilde{\boldsymbol{b}}_{\text{HD}}~|~\tilde{\boldsymbol{p}}_{\text{HD}}\right]$
and $\tilde{\boldsymbol{\alpha}}=\pi_{2}\left(\pi_{1}\left(\boldsymbol{\alpha}\right)\right)$,
note that $\tilde{\boldsymbol{b}}_{\text{HD}}$ corresponds to the
most-reliable independent positions of the received vector $\boldsymbol{r}$.

During the OSD algorithm, test-error patterns (TEPs) $\boldsymbol{e}_{i}$
of increasing weight are generated. They are added to the hard-decision
information bits $\tilde{\boldsymbol{b}}_{\text{HD}}$ on the MRB
and the corresponding codeword $\tilde{\boldsymbol{c}}_{i}$ is obtained
by re-encoding via the systematic generator matrix $\tilde{G}$. The
trivial TEP $\boldsymbol{e}_{0}=\boldsymbol{0}$ results in the order-0
OSD codeword $\tilde{\boldsymbol{c}}_{0}=(\tilde{\boldsymbol{b}}_{\text{HD}}+\boldsymbol{e}_{0})\cdot\tilde{G}=\tilde{\boldsymbol{b}}_{\text{HD}}\cdot\tilde{G}$.
The TEP $\boldsymbol{e}_{j}$ results in codeword $\tilde{\boldsymbol{c}}_{j}=\tilde{\boldsymbol{c}}_{0}+\boldsymbol{e}_{j}\cdot\tilde{G}$.
Undoing the permutations yields the estimate $\hat{\boldsymbol{c}}_{j}=\pi_{1}^{-1}\left(\pi_{2}^{-1}\left(\tilde{\boldsymbol{c}}_{j}\right)\right)$
of the original codeword $\boldsymbol{c}$.

After every re-encoding operation, the Euclidean distance between the
OSD codeword $\hat{\boldsymbol{c}}_{j}$ and the received vector $\boldsymbol{r}$
is calculated. If this distance is lower than that of the current
best OSD codeword, we select $\hat{\boldsymbol{c}}_{j}$ as the new
best codeword estimate. Note that for BPSK modulation, minimizing
the Euclidean distance is equivalent to minimizing the weighted Hamming
distance 
\[
\text{WHD}_j=\sum\limits _{\substack{1\leq i\leq n\\
\hat{c}_{j,i}\neq y_{i}
}
}\alpha_{i}.
\]
The OSD algorithm is terminated after a predetermined number of re-encodings.
For example, in OSD order 2, the following patterns are generated:

\begin{align*}
 & \text{weight 0}	&  	& \text{weight 1} 	&	& \text{weight 2}\\
 & 000...000 		&	& 000...001 		&	& 000...011\\
 &  				&  	& 000...010 		&  	& 000...101\\
 &  				&  	& 000...100 		&  	& 000...110\\
 & 					&  	& \vdots 			&  	& \vdots\\
 &  				&  	& 100...000			&	& 110...000
\end{align*}

It follows that in OSD order 1, $1+k$ patterns are generated, in
OSD order 2 we generate $1+k+\frac{1}{2}k(k-1)$ patterns, etc. Hence the
complexity of the algorithm is $O(k^{\text{OSD order}})$. In~\cite{Fossorier1995} it was shown that order-$l$ reprocessing is
asymptotically optimal (close to ML) for 
\[
l\ge\min\left\{ \left\lceil d_{H}/4-1\right\rceil ,k\right\},
\]
such that the complexity is determined by both $k$ and $d_{H}$.
Choosing the OSD order lower than the optimal allows a performance-complexity trade-off.
In this paper, in Section~\ref{sec_results}, we compare codes at the best possible OSD decoding performance.
\subsection{Improvements}
In the literature, several improvements to the original OSD algorithm
have been presented that aim to reduce the complexity
of the optimal decoder and offer a finer performance-complexity trade-off~
\cite{Fossorier2002,Valembois2004,Jin2006,Wu2007_1,Wu2007_2,Jin2007,Wu2008}.
In our implementation of the algorithm we used the probabilistic necessary
condition from~\cite{Wu2007_2}, the probabilistic sufficient
condition~\cite{Jin2006}, the reference re-encoding scheme~\cite{Wu2007_2},
the preprocessing rules from~\cite{Wu2007_1}, and the multiple biases
diversity scheme from~\cite{Jin2007}.

After having implemented these improvements, we can no longer use the rule 
$l\ge\min\left\{ \left\lceil d_{H}/4-1\right\rceil ,k\right\} $ to determine
the optimal OSD order. Furthermore, the parameters of the improvements also have to be set.
To determine if the decoder performs (near-)optimally, we make use of an ML lower bound 
calculated during computer simulation. Whenever the decoder outputs an erroneous estimate of the 
originally transmitted information word, the Euclidean distance between the original codeword $\boldsymbol{c}$
and the received vector $\boldsymbol{s}$ is evaluated. If this distance is larger than the
distance between the decoder output $\hat{\boldsymbol{c}}$ and $\boldsymbol{s}$, then the ML decoder
would also have made an erroneous decision.

\begin{figure}
\begin{center}
\includegraphics[angle=-90,width=1\columnwidth]{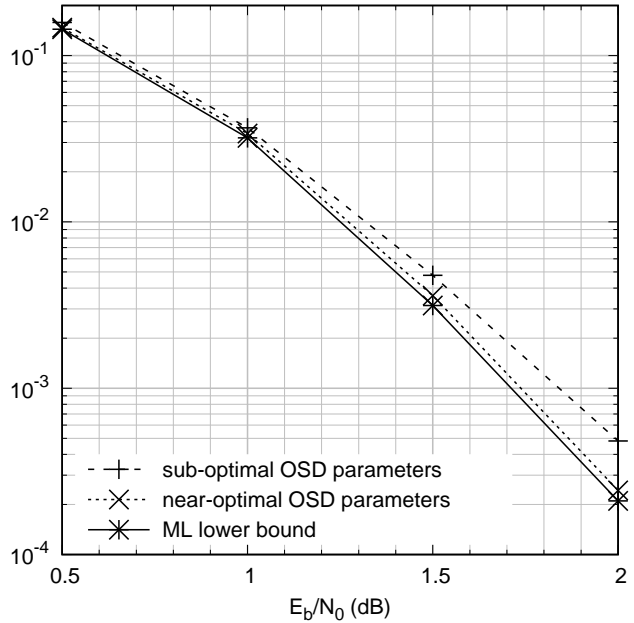}
\caption{\label{fig_plot_OSD_ML_lower} Word error rate versus signal-to-noise ratio for a (256,115) binary BCH code.
ML lower bound and word error rates for two sets of OSD parameters are shown.}
\end{center}
\end{figure}

Figure~\ref{fig_plot_OSD_ML_lower} shows the simulated performance of the (256,115)
extended BCH code for two different choices of OSD parameters. The ML lower bound derived from
the simulation with near-optimal parameters is also shown. We conclude that the second set
of parameters leads to near-optimal results for this particular code at these signal-to-noise (SNR) values.
The sub-optimal parameters, not considered in Section~\ref{sec_results}, 
are however suitable for some practical applications because they lead to
a decoding with a noticeably lower complexity at the expense of only a small loss of performance.

\section{List of Codes suited to Short-Length Error-Correction \label{sec_codes}}
We apologize for not considering convolutional codes and turbo codes 
(parallel concatenated conv. codes)~\cite{Berrou1996}, \cite{Benedetto1996}.
Results on turbo codes will be included in a future work.

Given the recent research activity in the Coding community \cite{Arikan2008,Tal2011,Mondelli2014,Vangala2015,Kudekar2015},
we had to consider Polar codes and Reed-Muller codes. Also, BCH codes are known to be good codes at short length~\cite{MacWilliams1977}, \cite{Blahut2003}. Finally, LDPC codes from modern coding theory~\cite{Richardson2008} are included. 
The list of binary codes regarded for performance comparison in the next section is:
\begin{itemize}
\item Reed-Muller codes: The code length is $n=2^{\ell}$. Take Arikan's kernel $G_2$~\cite{Arikan2008} 
and build its Kronecker product $\ell$ times, i.e. build $G_2^{\otimes \ell}$. 
Then, select the $k$ rows of largest Hamming weight to get the $k \times n$ generator matrix.
\item Polar codes: As for Reed-Muller codes, $k$ rows are selected from $G_2^{\otimes \ell}$.
These rows correspond to highest mutual information channels after $\ell$ splittings.
The generator matrix of the Polar code is found by exact splitting 
and adapted to each value of the channel parameter. 
For the BI-AWGN channel, we used density evolution \cite{Richardson2008} 
to split the channel and construct the code~\cite{Tanaka2009}.
\item BCH codes: Standard binary primitive $(n,k,t)$ BCH codes are built from their generator polynomial~\cite{MacWilliams1977}, \cite{Blahut2003}. An extension by one parity bit is made to get an even length. 
\item LDPC codes: Regular (3,6) low-density parity-check codes are built from a random bipartite Tanner graph~\cite{Richardson2008}.
Length-2 cycles are avoided, the number of length-4 cycles is reduced, but no other constraint was applied to the graph construction.
\end{itemize}

The use of a cyclic redundancy check (CRC) code to improve list decoding of polar codes 
was introduced by I.~Tal and A.~Vardy \cite{Tal2011}. Here, given the universal nature of Gaussian elimination
for ML decoding on the BEC and the universal nature of OSD decoding on the BI-AWGN channel,
the CRC code was jointly decoded with all of the codes listed above to investigate its influence on the performance.
By jointly we mean that a unique generator matrix is used for decoding.
This joint matrix is simply the product of the CRC matrix with the generator matrix of the original code $C$. 
Let $G$ be the $k \times n$ generator matrix of $C$. 
Let $G_{CRC}$ be the $(k-m)\times k$ generator matrix of the CRC code, where $m$ is the degree of the CRC polynomial.
Then, joint OSD decoding is based on the following generator matrix:
\[
G_{CRC} \times G.
\]
The serial concatenation has the CRC as outer code and the original error-correcting code $C$ as inner code.
It is clear that the CRC will scramble the original matrix $G$ making any code $C$ look like a random code.
We considered $m=16$ redundancy bits and the CRC-CCITT code with generator polynomial 
\[
g(x)=x^{16}+x^{12}+x^{5}+1.
\]

\section{Performance Results \label{sec_results}}
We ran computer simulations to  
obtain the performance of binary codes listed in the previous section.
Randomly generated data is transmitted using the systems described in sections~\ref{sec_bec}~and~\ref{sec_osd}.
At every considered value of $\epsilon$ for the BEC and $E_b/N_0$ for the BI-AWGN channel,
codewords were generated, transmitted, and decoded until 100 word errors occurred. During the
computer simulation on the BI-AWGN channel, the ML lower bound was also recorded but we omit it from the
figures to keep the graphs as clear as possible. The OSD parameters were chosen such that the
performance is near-ML and the ML lower bound (almost) coincides with the actual simulated performance
of the code.

Lower bounds on the optimal performance of finite-length codes exist for both the BEC and BI-AWGN channel.
On the figures we include for the BEC the Polyanskiy-Poor-Verd\'u (PPV) bound on the maximal achievable rate, 
from Theorem~53 in~\cite{Polyanskiy2010}, 
\[
P_{ew} \approx Q\left( \frac{1-\epsilon-R}{\sqrt{\epsilon (1-\epsilon)}} \sqrt{n} \right),
\]
where $P_{ew}$ is the word error probability at coding rate $R$, length $n$, and BEC parameter $\epsilon$. $Q(.)$ is the Gaussian-tail function.
Computer simulations below take $n=256$ and $R=1/2$ or close to $1/2$.
On the BI-AWGN channel, we include the word error probability of optimal spherical codes, by C.E.~Shannon~\cite{Shannon1959},
\begin{align*}
P_{ew} \approx & \frac{1}{\sqrt{n\pi}} \frac{1}{\sqrt{1+G^2} \sin \theta_0} \\
& \times \frac{\Big[G \sin \theta_0 \exp \Big(-\frac{RE_b}{N_0} + \frac{1}{2} \sqrt{\frac{2RE_b}{N_0}}G\cos \theta_0 \Big) \Big]^n}{\sqrt{\frac{2RE_b}{N_0}}G \sin^2 \theta_0 - \cos \theta_0},
\end{align*}
where $G = \frac{1}{2} \Big [ \sqrt{\frac{2RE_b}{N_0}} \cos \theta_0 + \sqrt{\frac{2RE_b}{N_0} \cos^2\theta_0 + 4} \Big ]$.
The cone half-angle $\theta_0$ is computed by solving
\[
2^{nR} \approx \frac{\sqrt{2\pi n} \sin \theta_0 \cos \theta_0}{\sin^n \theta_0}.
\]
The two above approximations from \cite{Shannon1959} are extremely accurate for lengths $n \ge 100$.\\

\begin{figure*}
\begin{center}
\includegraphics[angle=-90,width=0.65\paperwidth]{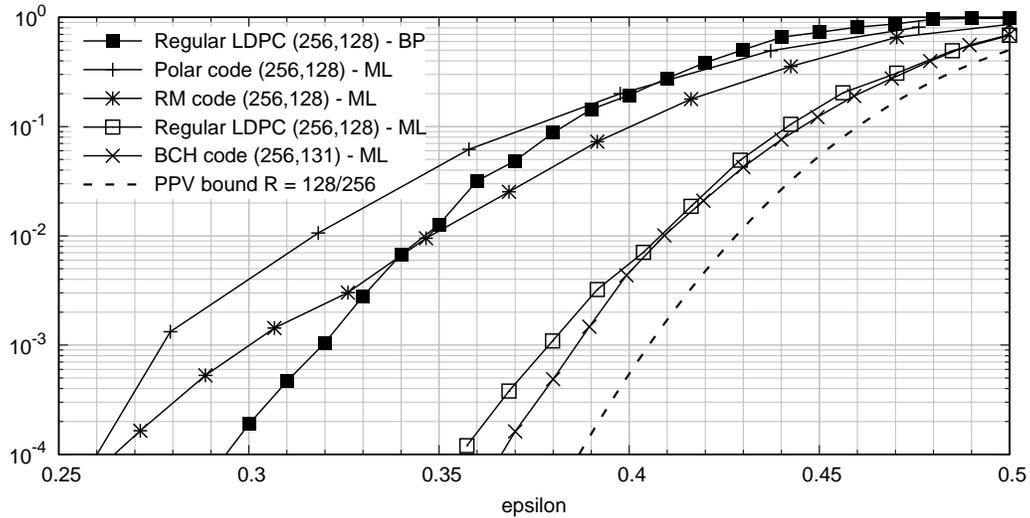}
\caption{\label{fig_plot_BEC_no_crc} Word error rate versus channel erasure probability. 
Performance comparison of codes at length 256 and rate $1/2$. No CRC. 
The RM code is weak at length $n=256$, however it behaves very close to LDPC and BCH at $n=128$.}
\end{center}
\end{figure*}

Figure~\ref{fig_plot_BEC_no_crc} shows the word error rate versus the channel erasure probability for LDPC, Polar, RM, and
BCH codes, all under ML decoding on the BEC. The performance of the LDPC code
under iterative belief-propagation (BP) decoding is also included. No CRC is used for this performance comparison. 
The binary (256,131) BCH code is outperforming all other codes. Notice that the regular-(3,6) binary LDPC code
has an excellent behavior.

\begin{figure*}
\begin{center}
\includegraphics[angle=-90,width=0.65\paperwidth]{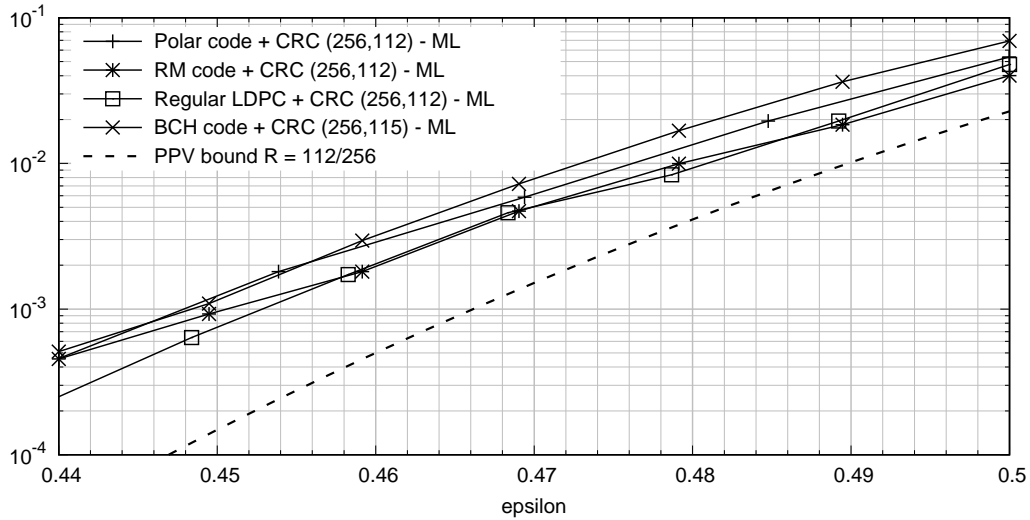}
\caption{\label{fig_plot_BEC_crc} Word error rate versus channel erasure probability. Performance comparison of codes with length 256 and rate $1/2$. A 16-bit CRC is concatenated with all codes. The horizontal scale is stretched to let us distinguish the small difference between codes.}
\end{center}
\end{figure*}

Figure~\ref{fig_plot_BEC_crc} shows the same codes on the BEC concatenated with the 16-bit CRC code given in Section~\ref{sec_codes}. The horizontal scale in Figure~\ref{fig_plot_BEC_crc} is from 0.44 to 0.5 only! 
All codes exhibit a performance within a small range of error rate. 
As mentioned in Section~\ref{sec_codes}, the CRC scrambles the original generator matrix and the universal
decoder (Gaussian elimination or OSD) does a joint decoding of both codes.

It is worth mentioning that the BCH code has dimension $k~=~115$, hence its coding rate is higher than other codes
and its PPV bound moves up with respect to the PPV bound at rate~$1/2$. This explains why the BCH code
appears to be weaker, in fact it is closer to its PPV bound than the other codes.

\begin{figure*}
\begin{center}
\includegraphics[angle=-90,width=0.65\paperwidth]{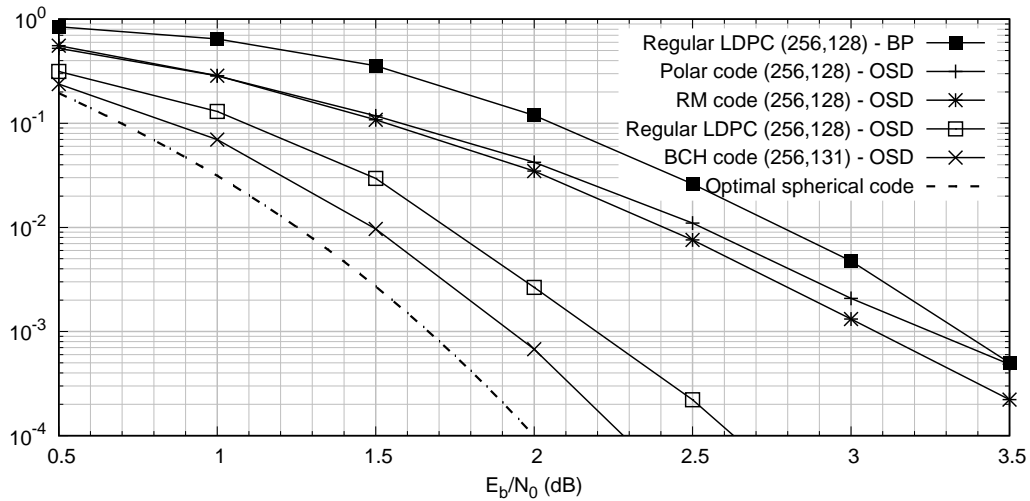}
\caption{\label{fig_plot_biawgn_no_crc}Word error rate versus signal-to-noise ratio. Performance comparison of codes with length 256 and rate $1/2$. No CRC.}
\end{center}
\end{figure*}

\begin{figure*}
\begin{center}
\includegraphics[angle=-90,width=0.65\paperwidth]{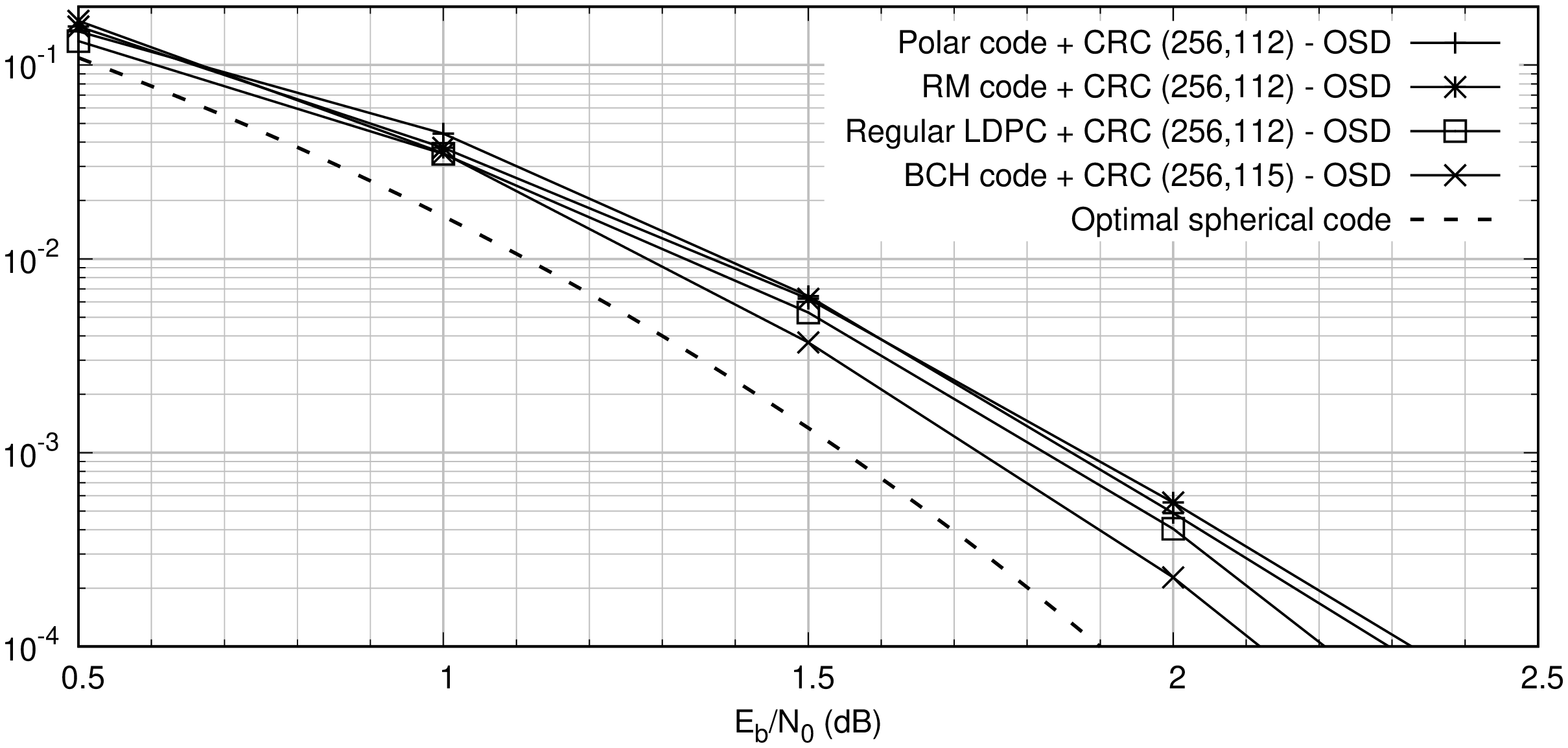}
\caption{\label{fig_plot_biawgn_crc}Word error rate versus signal-to-noise ratio. Performance comparison of codes with length 256 and rate $1/2$. A 16-bit CRC is concatenated with all codes. Codes with CRC cannot improve further due to the optimal spherical code bound (gap of 0.1-0.3 dB).}
\end{center}
\end{figure*}

The word error rate on the binary-input Gaussian channel is plotted on Figures~\ref{fig_plot_biawgn_no_crc} and~\ref{fig_plot_biawgn_crc}, without and with CRC respectively. Figure~\ref{fig_plot_biawgn_no_crc} also includes the LDPC
code under iterative belief-propagation (BP) decoding. With or without CRC, the (256,128) BCH code has the best performance versus the signal-to-noise ratio. As on the BEC, the CRC makes all codes behave almost like random codes, so the SNR gap between
the worst and the best codes is very small. 


\section{Conclusions \label{sec_conclusions}}
We have compared the performance of four different short-length linear binary codes on the binary erasure channel (BEC)
and the binary-input Gaussian (BI-AWGN) channel. The word error rate versus the channel parameter was plotted
for LDPC, Reed-Muller, Polar, and BCH codes.
In both channel scenarios, a universal, optimal$/$near-optimal decoder was used: 
the ML decoder for the BEC (via Gaussian elimination) and the OSD soft-decision decoder for the AWGN channel. 
From the computer simulation results, we conclude that the
BCH code outperforms Reed-Muller, Polar, and LDPC codes on both channels.
This behavior changes when we concatenate codes with a 16-bit CRC and perform joint decoding.
As a result, the performance curves of the different codes lie much closer together and the choice
of a good error-correcting code is not so critical. 

\section*{Acknowledgment}
Johannes~Van~Wonterghem would like to thank the Research Foundation in Flanders (FWO) for funding his PhD fellowship. 
The work of Joseph~J.~Boutros was supported by the Qatar National Research Fund (QNRF), 
a member of Qatar Foundation, under NPRP project 6-784-2-329. \\
The authors are grateful to Yury Polyanskiy for his precious and constructive comments on this work.


\end{document}